\documentclass[a4paper,11pt]{article}
\usepackage{pos}

\title{The role of Padé and D-Log Padé approximants in the context of the MUonE Experiment}
\ShortTitle{Convergence methods in the context of MUonE}

\author*[a]{Camilo Rojas P.}
\author[b]{Diogo Boito}
\author[b]{Cristiane Y. London}
\author[a]{Pere Masjuan}

\affiliation[a]{Grup de Física Teòrica, Departament de Física, Universitat Autònoma de Barcelona (UAB), and Institut de
Física d’Altes Energies (IFAE),\\ 
Campus UAB, E-08193 Bellaterra (Barcelona), Spain}

\affiliation[b]{Instituto de Física de São Carlos (IFSC), Universidade de São Paulo,\\
Caixa Postal 369, 13560-970 São Carlos, SP, Brazil}

\emailAdd{crojas@ifae.es}

\abstract{In the context of the anomalous magnetic moment of the muon, the hadronic contribution plays a crucial role, especially given its large contribution to the final error. Currently, lattice QCD simulations are in disagreement with dispersive calculations based on $e^+e^-$ hadronic cross sections. The new MUonE experiment intends to shed light on this situation extracting the hadronic contribution to the running of the electromagnetic coupling in the space-like region, $\Delta \alpha_{\rm had}(t)$, from elastic  $e\mu$ scattering. Still, due to the limited kinematic range that can be covered by the experiment, a powerful method must be devised to accurately extract the desired hadronic contribution from a new experiment of this type. In this work, we show how Padé and D-Log Padé approximants profiting from the analyticity of the correlator governing the hadronic contribution can be a powerful tool in reaching the required precision.}

\FullConference{%
  QNP2024 - The 10th International Conference on Quarks and Nuclear Physics,\\
  08-12 July 2024\\
  ICCUB, Barcelona, Spain}

\begin{document}
\maketitle

\section{Introduction}

The determination of the anomalous magnetic moment of the muon \(a_\mu = (g - 2)/2\) has seen significant progress through experimental results from FNAL E989 (2021, 2023) and BNL E821 (2006), achieving an impressive uncertainty of only 0.19 ppm~\cite{Muong-2:2021ojo,Muong-2:2023cdq,Muong-2:2006rrc}. However, the Standard Model (SM) prediction for \(a_\mu\), as recommended by the 2020 \(g-2\) Theory Initiative White Paper~\cite{Aoyama:2020ynm}, shows a deviation from the combined experimental results of 5.1\(\sigma\). This discrepancy primarily arises from the dispersive description of the Hadronic Vacuum Polarization (HVP) contribution to \(a_\mu\) (\(a_\mu^{\rm HVP}\)) using $e^+e^-$ data. Conversely, a lattice-based calculation for $a_\mu^{\rm HVP}$ reduces this tension to 2.0\(\sigma\)~\cite{Borsanyi:2020mff},
highlighting the need to resolve the differences between these two approaches. A recent work has shown that a hybrid lattice-dispersive strategy can improve even further the agreement with experiment \cite{boccaletti:2024high}.

Understanding the origin of the tension between dispersive and lattice-based determinations of \(a_\mu^{\rm HVP}\) is of crucial importance. In this context, the MUonE experiment has been designed to provide critical insight into this problem by measuring the hadronic contribution to the running of the electromagnetic coupling, \( \Delta \alpha_{\text{had}}(t) \), in the space-like region of elastic \(e\mu\) scattering~\cite{CarloniCalame:2015obs,Abbiendi:2016xup,Abbiendi:2677471}. MUonE is expected to cover approximately 86\% of the necessary integration range required to compute the leading-order contribution to the HVP, \(a_\mu^{\rm HVP,\,LO}\)~\cite{Abbiendi:2022oks,Ignatov:2023wma}.
However, a challenge arises in determining \(a_\mu^{\rm HVP,\,LO}\) using only MUonE data, as it requires extrapolation beyond the experimentally accessible region in the Euclidean \(t\)-variable, limited to the range \(-0.153~\text{GeV}^2 \leq t \leq -0.001~\text{GeV}^2\). To overcome this limitation, we propose the use of Padé approximants (PAs) and D-Log Padé approximants (D-Logs), which offer a systematic, simple, and model-independent approach to fit and extrapolate the data for the full determination of \(a_\mu^{\rm HVP,\,LO}\).

To demonstrate the effectiveness of our approach, we utilize a realistic model for the HVP function as presented in Ref.~\cite{Greynat:2022geu}, simulating the anticipated MUonE data. We apply both PAs and D-Logs to the exact Taylor expansion of the model function, and subsequently to pseudo-data generated from this model, to test the reliability and accuracy of our method.
In this work, we present only the results obtained from the canonical construction and the fitting to the simulated realistic data expected from the MUonE experiment. For a comprehensive and detailed description of the entire work, see reference~\cite{Boito:2024model}.

\section{Theoretical Framework}
The Standard Model (SM) calculation of \( a_\mu \) consists of four main contributions: Quantum Electrodynamics (QED), electroweak effects, HVP, and hadronic light-by-light scattering. The dominant uncertainty arises from the \( a^{\text{HVP, LO}}_\mu \) \cite{Aoyama:2020ynm}. The HVP can be computed using the polarization function which is related to the two-point correlator of the electromagnetic current, \( \Pi(q^2) \). 

Using a dispersive approach and the analytical properties of $\Pi(q^2)$ we can express $a_\mu^{\mathrm{HVP, \,LO}}$ as~\cite{Boito:2024model,Lautrup:1971jf}: 
\begin{equation}
    a_\mu^{\mathrm{HVP, \,LO}} = \dfrac{\alpha^2}{\pi} \int_0^1 \mathrm{d}x \,\,\, (1-x) \, \Delta\alpha_{\mathrm{had}}[t(x)],
    \label{eq:amu}
\end{equation}
where $\alpha$ is the electromagnetic fine-structure constant and we defined $\Delta\alpha_{\mathrm{had}}(t) = -4\pi\operatorname{Re}[\bar{\Pi}_{\mathrm{had}}(t)]$ as the hadronic contribution to the running of $\alpha$~\cite{Bernecker:2011gh}. Finally, $t$ is the space-like variable given by
\begin{equation}
    t = - \dfrac{x^2 \, m_\mu^2}{1-x}.
    \label{eq:ttox}
\end{equation}

The planned MUonE experiment is restricted to the interval \( x \in [0.2, 0.93] \), corresponding to \( -0.15~{\rm GeV}^2 \lesssim t \lesssim -0.001~{\rm GeV}^2 \), which is not sufficient for a full computation of \( a_\mu^{\rm HVP,\,LO} \). While extrapolation for \( 0 \leq x < 0.2 \) is relatively straightforward, as it covers a small interval in \( t \) near the origin, the extrapolation for \( 0.93 < x < 1 \), corresponding to \( 0.15~{\rm GeV}^2 \lesssim -t < \infty \), is significantly more complex and requires a careful uncertainty assessment.

\section{Padé and D-Log Padé approximants}


In this section, we give an overview of Padé Theory including PAs and D-Logs. The PAs are used to approximate a function $f(z)$ whose Taylor series is known and are given by a ratio of two polynomials, $Q_N(z)$ and $R_M(z)$ of orders $N$ and $M$ respectively\cite{Baker1996pade,Baker1975essentials} 

\begin{equation}
    \label{eq:pa}
    P_M^N(z) = \dfrac{Q_N(z)}{R_M(z)} = \dfrac{q_0 + q_1 \,z + \cdots + q_N \, z^N}{1 + r_1 \,z + \cdots + r_M \, z^M},
\end{equation}
where we adopted $r_0 = 1$. The PA parameters are determined by matching the Taylor coefficients of $f(z)$ to the expansion of the PA. Hence, the approximant $P_M^N(z)$ will reproduce the first $M+N+1$ coefficients of $f(z)$. The PAs approximate more efficiently the original function when compared to the Taylor series. They are also systematic and model independent and they can reproduce the analytic properties of the function. 

Since \( \Delta \alpha_{\text{had}}(t) \) may contain not only poles but also branch points across its full domain, we propose using a different type of approximants known as D-Logs. These approximants are specifically designed to handle functions with branch cuts. The goal is to approximate the logarithmic derivative of the original function, where isolated branch cuts turn into simple poles and more involved branch-cut structures can still be approximated by simple poles\footnote{Formally, these simple poles serve as a first approximation of the logarithmic derivative around $z=0$. When branch points are present in the original function, this approach also suggests a simple pole at the location of each singularity.}, and the residues are proportional to the multiplicity of each cut. By approximating this transformed function with a PA ($\bar{P}_M^N(z)$), we can subsequently reconstruct an approximation for the original function using the following D-Log ($D_M^N(z)$) definition~\cite{Baker1996pade}:

\begin{equation}
    \mathrm{Dlog}_M^N(z) \equiv D_M^N(z)= f(0) \exp{\left[\int \mathrm{d}z \, \bar{P}_M^N(z) \right]}, \label{eq:dlogpa}
\end{equation}
the factor $f(0)$ is a constant that reproduces the function’s value at $z=0$ when $f(0)\neq 0$.\footnote{This constant is included because, after taking the derivative this factor disappears.}
The $D_M^N$  then reproduces exactly the first $M+N+2$ coefficients of $f(z)$ and can be used to predict the $(M+N+3)$-th coefficient and higher. The D-Logs can provide an unbiased estimate of the location of the branch point and also the multiplicity of the cut, since no assumptions about its position or its multiplicity are made~\cite{Baker1996pade,Boito:2018rwt,Boito:2021scm}.

With these methods, the convergence of some sequences of PAs to the original function is guaranteed for specific types of functions, for example, Stieltjes functions, which is the case of \( \Delta \alpha_{\text{had}}(t) \)~\cite{Masjuan:2009wy,Aubin:2012me}. A Stieltjes function is a function that can be represented by the following Stieltjes integral
\begin{equation}\label{eq:stieltjes}
    f(z) = \int_0^\infty \dfrac{\mathrm{d}\phi(u)}{1+zu}\,,
\end{equation}
where $\phi(u)$ is a bounded non-decreasing function on the interval $0 \leq u < \infty$.
In the case of Stieltjes functions, Padé Theory assures that the poles of approximants of the type $P_M^{M+ k}$ with $k\geq-1$ to the function $\Delta\alpha_{\mathrm{had}}$ are always real and positive~\cite{Baker1975essentials,Baker1996pade}. By extension, whenever the D-Logs are built to a Stieltjes function and the associated PA can be assured Stieltjes, the branch point of D-Logs of the type $D_N^N$, $D_N^{N+1}$ and $D^N_{N+1}$ to $\Delta\alpha_{\mathrm{had}}$ are also real and positive. Furthermore, since $\Delta\alpha_{\mathrm{had}}$ scales as ${\cal O}(t^1)$ for small $t$, the PA sequences $P_N^N$ and both $P_N^{N+1}$ and $P^N_{N+1}$ bound the original function. In our case, the fastest convergence is obtained with the $P_N^{N+1}$, then the convergence theorem for Stieltjes functions reads
\begin{equation}
    P_1^1(t) \leq P_2^2(t) \leq \dots \leq \Delta\alpha_{\mathrm{had}}  \leq \dots  \leq P_2^3(t)\leq P_1^2(t) \, .
    \label{eq:convergence}
\end{equation}
For the D-Log case, a similar pattern is found \cite{Boito:2024model}. The sequence $D_N^N$ together with $D^N_{N+1}$ and $D_N^{N+1}$ bound the original function. The observed convergence pattern is as follows
\begin{equation}
    D_1^1(t) \leq D_2^2(t) \leq \dots \leq \Delta\alpha_{\mathrm{had}} \leq \dots \leq D_3^2(t) \leq D_2^1(t)\,.
    \label{eq:convergenceDlogs}
\end{equation}
In our study, we found the fastest convergence with the \( D^N_{N+1} \) sequence (as stated in Eq.~(\ref{eq:convergenceDlogs})), which closely approximates the model used for \( \Delta \alpha_{\mathrm{had}} \) introduced in the next section.

\section{A Model for the Euclidean Correlator}

To test the effectiveness of our method, we generate toy data sets using a phenomenological model for the hadronic correlator \( \operatorname{Im} \Pi_{\mathrm{had}}(s) \), introduced by Greynat and de Rafael~\cite{Greynat:2022geu} (GdR model). This model, inspired by chiral perturbation theory and perturbative QCD, allows us to compute \( \Delta \alpha_{\mathrm{had}} \) through a dispersion relation. 
We consider it to be realistic because it produces a representation of \( \Delta \alpha_{\mathrm{had}} \) as a Stieltjes function, consistent with QCD expectations. By computing the integral of the model, we obtain a reference value of \( a_\mu^{\mathrm{HVP,\,LO}} \) as \( 6992.4 \times 10^{-11} \). This value will serve as a benchmark for comparison with the results obtained using PAs and D-Logs in the rest of the analysis, helping to assess the accuracy and reliability of these methods.

\begin{figure}[!h]
    \centering \includegraphics[width=0.45\textwidth]{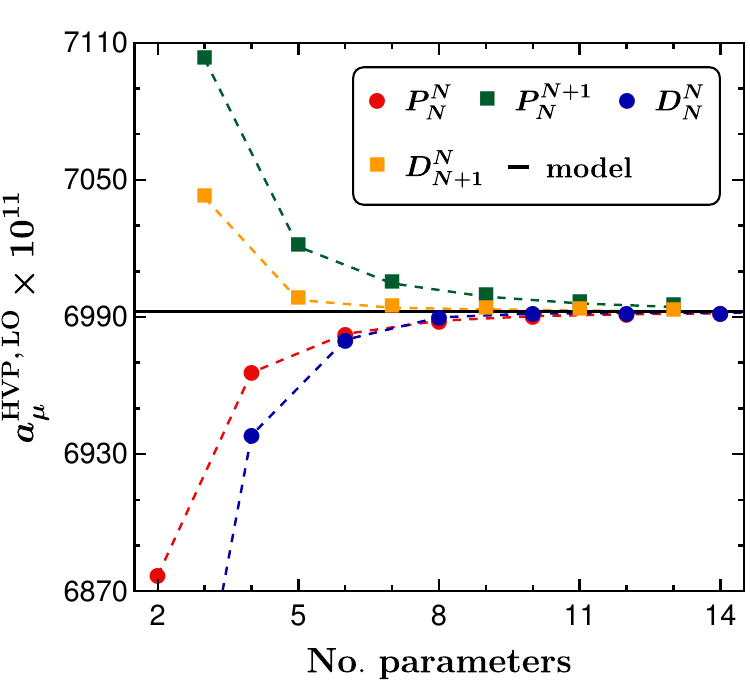}
    \caption{Comparison between PA and D-Log estimates of $a_\mu^{\mathrm{HVP,\,LO}}$ and the value predicted by the model of Greynat and de Rafael (black solid line). `No. parameters' refers to the value of $N+M$ used for each approximant.}
    \label{fig:amutaylor}
\end{figure}

For the canonical construction of PAs and D-Logs applied to the Taylor expansion of this model, we observe a convergence pattern that satisfies the PAs theorems for Stieltjes functions. Once the PAs and D-Logs are constructed, we can proceed to estimate the value of \( a_\mu^{\mathrm{HVP,\,LO}} \), as shown in Figure \ref{fig:amutaylor}.


\section{Methodology}
The methodology uses sequences of PAs and D-Logs to systematically fit toy data generated for the MUonE experiment by minimizing a modified $\chi^2$ function. These fitting functions are constructed from a general power series representing \( \Delta \alpha_{\text{had}}(t) \), and constraints based on Stieltjes function properties are applied to ensure real poles, zeros and cuts on the positive axis. This approach enables reliable extrapolations beyond the experimental region while providing systematic uncertainty estimates. PAs used as fitting functions were already applied in similar contexts and the convergence theorems, which are not strictly valid when the approximants are not built from the Taylor series, are apparently satisfied in all these cases~\cite{boito:2023probing,Gonzales:2021pade,Escribano:2015yup, Escribano:2013kba,Masjuan:2008fv}, a closer work in the context of the $g_{\mu}-2$ also use PAs as fitting functions~\cite{Masjuan:2017tvw}.

\section{Results}

Here, we present only the most relevant results, using the approximants to fit pseudo-data with realistic errors by incorporating fluctuations and uncertainties expected in the MUonE experiment. For the complete analysis, we refer to Ref.~\cite{Boito:2024model}. We generated 1000 toy data sets for the function \( \alpha \, \Delta\alpha_{\mathrm{had}}(x) \times 10^5 \), each consisting of 30 data points equally distributed in the region $x \in [0.2,0.93]$ and corresponding to expected MUonE bin sizes. The data sets were generated based on a Gaussian distribution around the value predicted by the GdR model with an error ranging from 0.7\% for larger values of \(x\) to 6.7\% for \(x \approx 0.2\).\footnote{Thanks to Giovanni Abbiendi, Carlo Carloni Calame, and Graziano Venanzoni for providing us with the values of the expected uncertainties of the MUonE experiment.} Additionally, the \( a_\mu^{\mathrm{HVP,\,LO}} \) of each data set was calculated. The median and 68\% confidence level (CL) of this distribution yield \( a_\mu^{\mathrm{HVP,\,LO}} = (6991^{+22}_{-20}) \times 10^{-11} \), which serves as a benchmark for comparison with predictions from PA and D-Log approximants.


After and during the fitting process to each toy data set, we inspected the approximants for the appearance of defects, which occur due to cancellations between poles and zeros in the PAs or in the D-Logs when specific conditions are met (e.g., merging branch points or exponents becoming zero). Approximately 30\% of the \(P_2^2\) fits and 56\% of the \(P_2^3\) fits were discarded due to these defects, while for D-Logs, the rates were lower, with 0.7\% discarded for \(D_2^3\) and 4\% for \(D_3^3\). We found that approximants with more than six fit parameters were prone to large uncertainties, so only lower-order approximants were used.

\begin{table}[t!]
    \begin{center}
        \caption{Results for $a_\mu^{\mathrm{HVP,\,LO}}$ from the PAs and D-Logs used as fitting functions to toy data sets together with the final values for $\chi^2/{n_{\mathrm{dof}}}$ of the respective approximants. Final results for both methods are also presented.
        }
        \begin{tabular}{ccc|ccc}
            \toprule
            & $a_\mu^{\mathrm{HVP,\,LO}} \times 10^{11}$ & $\chi^2/{n_{\mathrm{dof}}}$ & & $a_\mu^{\mathrm{HVP,\,LO}} \times 10^{11}$ & $\chi^2/{n_{\mathrm{dof}}}$ \\ \hline
	   $P_1^1$ & $6938 \pm 21$ & $1.01_{-0.25}^{+0.27}$ & $D_2^1$ & $7052^{+66}_{-71}$ & $1.01_{-0.26}^{+0.26}$ \\[0.1cm]
	   $P_1^2$ & $7042^{+114}_{-104}$ & $1.01_{-0.26}^{+0.28}$ & $D_2^2$ & $6956^{+96}_{-65}$ & $1.05_{-0.27}^{+0.28}$ \\[0.1cm]
	   $P_2^2$ & $6980^{+46}_{-34}$ & $1.05_{-0.27}^{+0.29}$ & $D_3^2$ & $6999^{+48}_{-39}$ & $1.10_{-0.28}^{+0.29}$ \\[0.1cm]
	   $P_2^3$ & $6994^{+85}_{-49}$ & $1.11_{-0.31}^{+0.29}$ & $D_3^3$ & $6977^{+72}_{-53}$ & $1.14_{-0.29}^{+0.30}$ \\ \hline
  Final result & $6987^{+46}_{-34}$ & --- & Final result & $6988^{+48}_{-39}$ & --- \\     \bottomrule
            \label{tab:PAsamu}
        \end{tabular}
    \end{center}
\end{table}

The final value for \( a_\mu^{\mathrm{HVP,\,LO}} \) is derived from the median of accepted fits, with uncertainties at a 68\% CL. Both PAs and D-Logs results show good agreement with the model prediction (see Tab.~\ref{tab:PAsamu}), and because the errors are primarily driven by the extrapolation beyond \(x = 0.93\) up to $x=1$, by reducing the extrapolation to \(x=0.990\) decreases the uncertainty by 25\%, covering 99.1\% of the \( a_\mu^{\mathrm{HVP,\,LO}} \) value. The final systematic uncertainties are small and do not significantly change with the extrapolation limit. 

\section{Conclusion}


We demonstrate the effectiveness of PAs and D-Logs as model-independent tools for fitting and extrapolating toy data that simulates the expected results of the MUonE experiment, with the aim of extracting \( a_\mu^{\mathrm{HVP,\,LO}} \). These approximants take advantage of the known properties of the hadronic contribution to the running of the fine-structure constant, which is a Stieltjes function in the $t$-variable~\cite{Aubin:2012me}, thus benefiting from well-established convergence theorems for PAs, even though a departure from the conditions where the theorems are strictly valid is inevitable when dealing with experimental data.

We observed that different sequences of approximants bound the true value of \( a_\mu^{\mathrm{HVP,\,LO}} \), allowing us to estimate the final result by averaging the highest-order approximants, \( P_2^2 \), \( P_2^3 \), \( D_3^2 \), and \( D_3^3 \). Our final estimates were in excellent agreement with the expected model values, with deviations of less than 0.06\% for PAs and 0.05\% for D-Logs. The uncertainty was dominated by the extrapolation beyond the data region, but this could be reduced by limiting the extrapolation range (e.g., up to \( x_{\rm max}=0.990 \), covering 99.1\% of the integrand).

Our explorations show that this method is superior to the use of a single, fixed, fitting function, which may carry a model dependence and an associated systematic uncertainty that would be difficult to estimate on the basis of real experimental data. Our method which takes fully into account the analytic properties of the original function provides a reliable, model-independent framework for the extraction of \( a_\mu^{\mathrm{HVP,\,LO}} \) with competitive, though conservative, uncertainty.

\section*{Acknowledgements}
The work of CYL was financed by the S\~ao Paulo Research Foundation (FAPESP) grant No.~2020/15532-1 and 2022/02328-2.
DB’s work was supported by FAPESP grant No.~2021/06756-6 and by CNPq grant No.~308979/2021-4. The work of PM has been supported by the Ministerio de Ciencia e Innovación under grant PID2020-112965GB-I00 and by the Secretaria d’Universitats i Recerca del Departament d’Empresa i Coneixement de la Generalitat de Catalunya under grant 2021 SGR 00649. IFAE is partially funded by the CERCA program of the Generalitat de Catalunya. CR's work has been supported with funding from Colfuturo and Colciencias of Colombia.

\bibliographystyle{jhep}
\bibliography{skeleton}



\end{document}